\documentclass[a4paper]{article}

\usepackage{ISCSLP2024}
\usepackage[T1]{fontenc}
\usepackage{graphicx}
\usepackage{multirow}
\usepackage{multirow}
\usepackage{makecell}
\usepackage{enumitem}
\usepackage{bm}
\usepackage{adjustbox}
\usepackage{bigstrut,multirow,rotating,booktabs}
\usepackage[bottom]{footmisc} 

\begin{document}

\title{Leveraging Prompt Learning and Pause Encoding for Alzheimer’s Disease Detection\vspace{-0.1em}\thanks{* corresponding author. This work is supported in part by National Social Science Foundation of China (23AYY012). This research was supported by the Supercomputing Center of the USTC.}}
\name{Yin-Long Liu$^1$, Rui Feng$^1$, Jia-Hong Yuan$^{1,2*}$, Zhen-Hua Ling$^{1,2}$}
\address{
  $^1$National Engineering Research Center of Speech and Language Information Processing, \\ University of Science and Technology of China, Hefei, P. R. China\\
  $^2$Interdisciplinary Research Center for Linguistic Sciences, \\ University of Science and Technology of China, Hefei, P. R. China}
\email{\{lyl2001, fengruimse\}@mail.ustc.edu.cn, \{jiahongyuan, zhling\}@ustc.edu.cn}

\maketitle

\begin{abstract}
Compared to other clinical screening techniques, speech-and-language-based automated Alzheimer’s disease (AD) detection methods are characterized by their non-invasiveness, cost-effectiveness, and convenience. Previous studies have demonstrated the efficacy of fine-tuning pre-trained language models (PLMs) for AD detection. However, the objective of this traditional fine-tuning method, which involves inputting only transcripts, is inconsistent with the masked language modeling (MLM) task used during the pre-training phase of PLMs. In this paper, we investigate prompt-based fine-tuning of PLMs, converting the classification task into a MLM task by inserting prompt templates into the transcript inputs. We also explore the impact of incorporating pause information from forced alignment into manual transcripts. Additionally, we compare the performance of various automatic speech recognition (ASR) models and select the Whisper model to generate ASR-based transcripts for comparison with manual transcripts. Furthermore, majority voting and ensemble techniques are applied across different PLMs (BERT and RoBERTa) using different random seeds. Ultimately, we obtain maximum detection accuracy of 95.8\% (with mean 87.9\%, std 3.3\%) using manual transcripts, achieving state-of-the-art performance for AD detection using only transcripts on the ADReSS test set.
\end{abstract}
\vspace{-1mm}
\section{Introduction}
\vspace{-1mm}
Alzheimer’s disease (AD), the most common cause of dementia, is a neurodegenerative disease that worsens over time and causes irreversible damage to the brain, manifested by a persistent deterioration of an individual’s cognitive and functional abilities, including language, memory, attention, and executive function \cite{nestor2004advances}. Studies have shown that mid-course intervention before neuronal degeneration in the brain can effectively alleviate the problem, so early detection of AD is crucial \cite{nestor2004advances,de2019brief}. In contrast to traditional clinical detection methods, speech-and-language-based automatic AD diagnosis technology has become a research hotspot due to its non-invasive, low-cost and convenient characteristics. Recent ADReSS challenge has also promoted the development of related technologies \cite{luz2020fuente}.
\vspace{-1mm}
\subsection{Related studies of speech and language in AD detection}
\vspace{-1mm}
Current research primarily utilizes two types of features extracted from spontaneous speech: acoustic features from raw audio signals and linguistic features from transcripts \cite{luz2021detecting,pan2021using}. Both features offer unique insights into cognitive decline associated with AD. Based on the recent studies \cite{liu2024clever,mei2023ustc,syed2020automated,syed2021tackling,rohanian2021alzheimer}, acoustic features used in AD detection can be divided into prosody, duration of pauses, vocal quality, emotional embeddings, pre-trained models embeddings features, and more. Key linguistic features include, but are not limited to, lexical richness, syntactic complexity and pre-trained textual embedding features \cite{martinc2020tackling,qiao2021alzheimer,santos2017enriching}. Previous research has demonstrated that, compared to using acoustic features alone, employing linguistic features extracted from participants' speech transcripts is more effective in distinguishing between AD and non-AD \cite{syed2020automated,li2021comparative}. Advanced natural language processing (NLP) techniques, particularly those involving pre-trained language models (PLMs) like BERT, have greatly improved the analysis of linguistic features. 
PLMs are trained on vast amounts of text data. They can capture complex patterns and contextual information that are critical for identifying cognitive decline. 
Wang et al. \cite{wang2022exploring} investigated the use of feature and model combination approaches to improve the robustness of domain fine-tuning of BERT and RoBERTa pre-trained text encoders, obtaining an AD detection accuracy of 91.67\% on manual transcripts of the ADReSS test set. 
\vspace{-1mm}
\subsection{The application of PLMs in AD detection}
\vspace{-1mm}
\label{sec:application of PLMs}
Currently, the use of PLMs in AD detection typically involves two approaches, the first is utilizing them as feature extractors, and the second is fine-tuning (refers to the traditional fine-tuning mentioned in the following sections) them \cite{li2021comparative,wang2022exploring,zhu2021wavbert,wang2023exploiting}. While using PLMs as feature extractors can leverage pre-trained representations efficiently, this approach is often limited by its inability to adapt to domain-specific nuances, and lack of task-specific optimization. The second approach can overcome these limitations. However, the objective during the fine-tuning phase is inconsistent with the objectives of the pre-training phase of PLMs.
PLMs like BERT primarily learn during the pre-training phase through two tasks: masked language modeling (MLM) and next sentence prediction (NSP) \cite{devlin2018bert,liu2019roberta}. The objectives of the two tasks are inconsistent with that of the fine-tuning method aimed at AD binary classification mentioned above. This inconsistency  may result in the PLMs overlooking fine-grained information, such as semantic coherence and pragmatic features, which are valuable for AD detection during the fine-tuning process.
\vspace{-1mm}
\subsection{The work of this paper}
\vspace{-1mm}
To address the inconsistency issue of objectives between the traditional fine-tuning (TFT) phase and the pre-training phase of PLMs for AD detection, as mentioned in Section \ref{sec:application of PLMs}, we investigated a novel fine-tuning paradigm based on prompts. By inserting prompt templates into the transcript inputs, we converted the text classification task into an MLM task, thereby achieving better consistency with the pre-training objectives. Previous studies have demonstrated the effectiveness of prompt-based fine-tuning (PBFT) in tasks such as text classification and text matching \cite{zhu2023prompt,feng2024prompt,xu2022match}, but there has been limited research in the medical domain \cite{zheng2024exploring}, particularly in the field of AD detection. In this paper, we converted the problem of classifying AD and non-AD labels into a word probability prediction task. Specifically, we insert a prompt template into the input text: ``The diagnosis result is [MASK]”. After fine-tuning the PLMs with this prompt on the training data, the models are required to predict whether the ``[MASK]” token is the word “alzheimer or “healthy” . This prompt-based approach provides the model with a clear prediction framework, which is consistent with the pre-training tasks. Consequently, it leverages the semantic and contextual knowledge acquired during pre-training phase more effectively and reduces the extent of model parameter adjustments required during fine-tuning phase. Different positions to insert the prompt template were also explored. To compare with the aforementioned PBFT method, we also implemented the TFT method, which only inputs the transcripts and does not include the aforementioned prompt phrases. 
Additionally, we implemented pause encoding using the timestamp outputs from forced alignment. To the best of our knowledge, no previous research has combined PBFT with pause encoding for AD detection. Recent advancements in automatic speech recognition (ASR) technologies, such as wav2vec 2.0 \cite{baevski2020wav2vec} and Whisper \cite{radford2023robust}, have significantly enhanced the viability of ASR transcripts for AD classification. 
To compare with the manual transcripts, we investigated the impact of ASR transcripts on AD classification performance. Specifically, we compared different ASR models and selected the Whisper model
for further experiments. Each sample in the ADReSS dataset includes dialogue content between the interviewer and the subject, and we explored the effect of the interviewer's transcripts on AD detection. To mitigate the overfitting risk of PLMs on small datasets and enhance robustness, mjority voting and ensemble techniques were further applied across BERT and RoBERTa using different random seeds. The main contributions of this paper are summarized below:
\begin{itemize}
\item It presents the first work combining PBFT with pause encoding for AD detection. Experimental results show that PBFT generally outperforms the TFT paradigm and confirm the effectiveness of pause encoding.
\item It compares the performance of various ASR models on the ADReSS dataset and investigates the impact of ASR-based versus manual transcripts on AD detection performance.
\item It achieves a state-of-the-art (SOTA) AD detection accuracy of 95.8\% using only manual transcripts on the ADReSS test set.
\end{itemize}

The rest of this paper is organized as follows. Section \ref{sec:Data} presents
the data and transcripts generation process. 
Section \ref{sec:Methods} describes two methods for fine-tuning PLMs and pause encoding. Section \ref{sec:Experiments} reports the experimental procedures, results, and related analysis. Finally, conclusions are drawn and future work is discussed in Section \ref{sec:Conclusions}.
\vspace{-1mm}
\section{Data}
\vspace{-1mm}
\label{sec:Data}
In this paper, the ADReSS challenge dataset from INTERSPEECH 2020 \cite{luz2020fuente} is used for training and evaluating the AD detection system. 
It is selected from the Pitt Corpus in the DementiaBank database \cite{becker1994natural}. 
The data consists of speech recordings and corresponding manual transcripts of spoken picture descriptions elicited from subjects and guidance phrases from the interviewers. 
Each sample has a binary AD label (alzheimer or healthy).
The training set includes 108 subjects 
and the test set includes 48 subjects.  
both balanced in terms of gender, age and binary labels. 
The data are in English and undergo acoustic enhancement through noise removal and audio volume normalization. 
Further details of the dataset are described in \cite{luz2020fuente}.

The following subsections describe how we processed the ADReSS dataset to obtain the  manual and ASR transcripts.
\vspace{-1mm}
\subsection{Manual Transcripts}
\vspace{-1mm}
The transcripts in the ADReSS dataset were annotated using the CHAT (Codes for the Human Analysis of Transcripts) format \cite{macwhinney2014childes}. 
This format standardizes the transcripts of spoken language, including punctuation, speaker identification, and the notation of nonverbal sounds and actions. 
We processed these annotations to ensure that the manual transcripts accurately correspond to what the participants actually said. Behavioral noises such as “\&=laughs” and “\&=coughs” were removed. Symbols such as \&, @, (.), (..), (...), <, >, /, and xxx were also removed. 
Instances of “word [x n]” were replaced by repeating “word” n times. And then, we obtained manual transcripts that represent what were actually produced in speech (including guidance phrases from the interviewers). Additionally, based on the annotations, we also obtained transcripts containing only the subjects’ speech.

\setcounter{footnote}{0}
\vspace{-1mm}
\subsection{ASR Transcripts}
\vspace{-1mm}
We compared two advanced ASR models, wav2vec 2.0 \cite{baevski2020wav2vec} and Whisper \cite{radford2023robust}, to determine which model could achieve more accurate transcription results on the pathological speech dataset. For the wav2vec 2.0 model, we utilized three fine-tuned versions: wav2vec2-large-960h\footnote {https://huggingface.co/facebook/wav2vec2-large-960h}, wav2vec2-large-960h-lv60-self\footnote{https://huggingface.co/facebook/wav2vec2-large-960h-lv60-self}, and wav2vec2-large-xlsr-53-english\footnote{https://huggingface.co/jonatasgrosman/wav2vec2-large-xlsr-53-english}. For the Whisper model, whisper-large-v3\footnote{https://huggingface.co/openai/whisper-large-v3} is utilized. The ASR-based transcripts include speech from both the subjects and the interviewers. All these models are available on HuggingFace.
\vspace{-1mm}
\section{Methods}
\label{sec:Methods}
In this paper, we used BERT and RoBERTa as the PLMs to be fine-tuned. Experiments were conducted on base-sized\footnote{https://huggingface.co/google-bert/bert-base-uncased}\footnote{https://huggingface.co/FacebookAI/roberta-base} or large-sized\footnote{https://huggingface.co/google-bert/bert-large-uncased}\footnote{https://huggingface.co/FacebookAI/roberta-large} versions.

\subsection{Traditional Fine-Tuning (TFT)}
\label{sec:Traditional Fine-Tuning}
\begin{figure}[t]
    \centering
\includegraphics[width=0.75\linewidth]{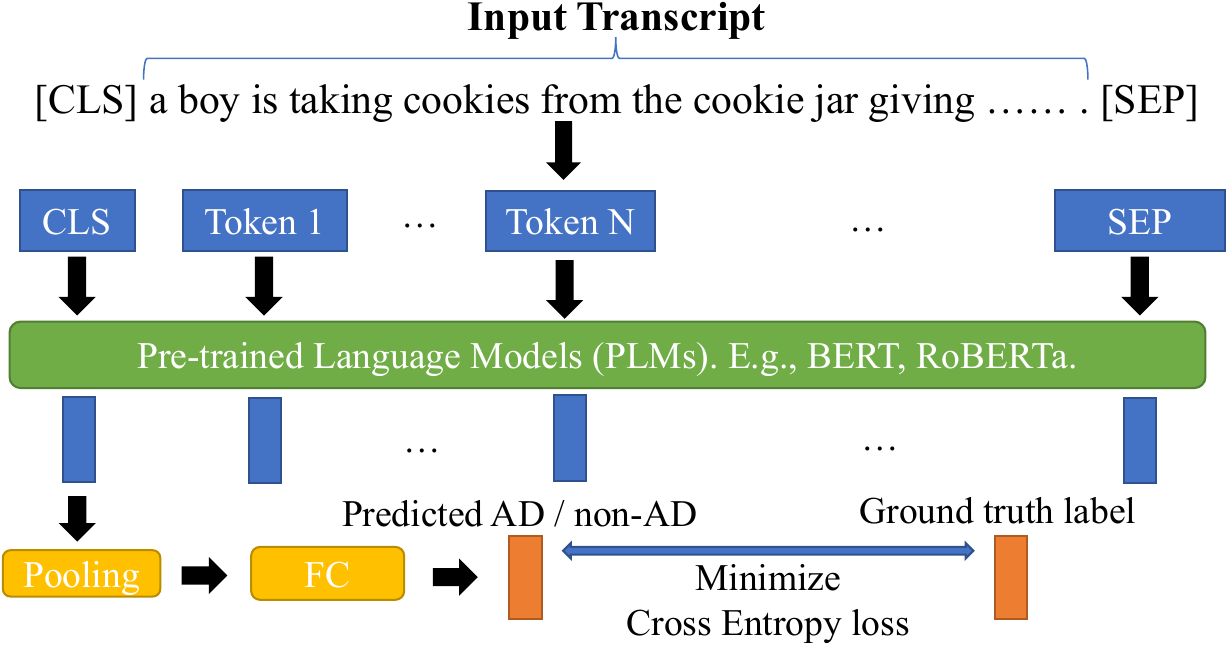}
    \caption{Overall structure for TFT.}
    \label{fig:traditional_fine_tuning}
     \vspace{-0.7cm}
\end{figure}

TFT method refers to using the “BertForSequenceClassification” or “RobertaForSequenceClassification” classes from the Transformers library\footnote{https://github.com/huggingface/transformers}, as illustrated in Figure \ref{fig:traditional_fine_tuning}. The input text is tokenized into tokens, including special tokens [CLS] and [SEP]. Each token, is converted into embeddings using embedding layers and then passed through transformer layers. The final hidden state of the [CLS] token, which captures the overall meaning of the input, is then passed to a sequence classifcation head (a linear layer with the sigmoid activation function over the average pooled output) to produce the classification logits and then determine whether the label is AD or non-AD.
\vspace{-1mm}
\subsection{Prompt-based Fine-Tuning (PBFT)}
\begin{figure}[t!]
    \centering
\includegraphics[width=0.75\linewidth]{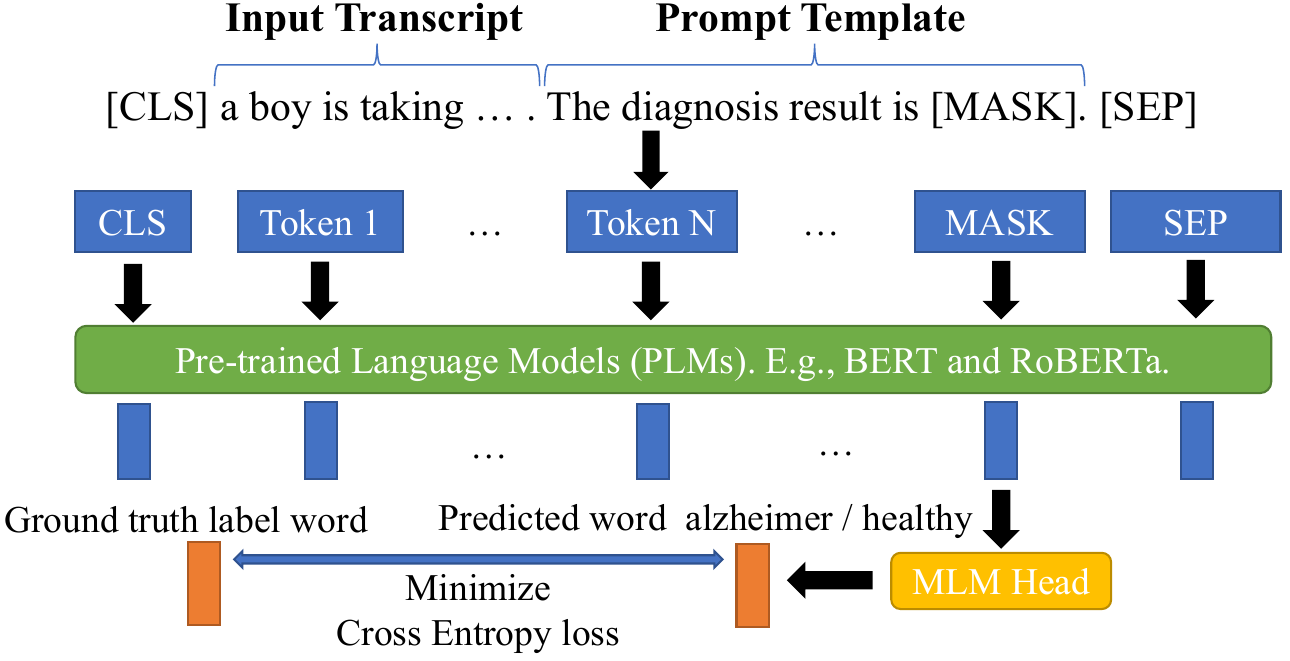}
    \caption{Overall structure for PBFT.}
    \label{fig:prompt_learning}
\vspace{-0.7cm}
\end{figure}

\vspace{-1mm}
The PBFT converts the text classification task into an MLM task by using manually designed prompt templates. Therefore, the model now needs to predict the probability of the label word to fill in the masked position, as illustrated in Figure \ref{fig:prompt_learning}. For this prompt template “The diagnosis result is [MASK]”, the “[MASK]” is the word “alzheimer” or “healthy”, corresponding to the AD or non-AD label, respectively. Each label word corresponds to one token in the vocabulary of BERT or RoBERTa. Unlike the TFT method mentioned in Section \ref{sec:Traditional Fine-Tuning}, the transcripts need to be concatenated with the prompt template before being input into the PLMs, and the final hidden state representation of the “[MASK]” 
 token is used to pass into MLM Head to predict logits
representing the probablities for the AD and non-AD label words to
fill in the masked position given the corresponding vocabulary. The logits are then normalized using the softmax function to compute the cross entropy loss for the label words “alzheimer” and “healthy”. We experimented with inserting the prompt template in two different positions (before or after the transcript) to investigate the impact of template positions on the correct prediction of label words, namely: “The diagnosis result is [MASK] +transcript” or “transcript + The diagnosis result is [MASK]”. We used the OpenPrompt framework \cite{ding2021openprompt} based on Pytorch to implement the PBFT with PLMs.

\begin{table}[b]
\vspace{-0.5cm}
    \caption{Mean WER (\%) of the ASR models on ADReSS.}
    \label{tab:WER of the ASR models}
    \centering
\resizebox{1.0\linewidth}{!}{
\begin{tabular}{cccccc}
\toprule
ASR model                      & \begin{tabular}[c]{@{}c@{}}All \end{tabular} & \begin{tabular}[c]{@{}c@{}}Healthy \end{tabular} & \begin{tabular}[c]{@{}c@{}}Alzheimer \end{tabular} & \begin{tabular}[c]{@{}c@{}}Training set \end{tabular} & \begin{tabular}[c]{@{}c@{}}Test set \end{tabular} \\
\midrule
wav2vec2-large-960h            & 50.19                                            & 42.68                                            & 57.78                                             & 49.74                                                     & 51.19                                                \\
\midrule
wav2vec2-large-960h-lv60-self  & 43.19                                            & 35.97                                           & 50.48                                           & 42.44                                                       & 44.84                                                \\
\midrule
wav2vec2-large-xlsr-53-english & 64.85                                             & 57.03                                          & 72.74                                          & 65.49                                                   & 63.43                                     \\
\midrule
\textbf{whisper-large-v3}& \textbf{34.24}& \textbf{28.47}& \textbf{40.07}& \textbf{32.97}& \textbf{37.06}\\
 \bottomrule
\end{tabular}}
\end{table}
\vspace{-2mm}
\subsection{Pause Encoding}
\vspace{-1mm}
We encoded the pauses between words in the input transcripts to better capture disfluency information. The specific methodology is as follows. The processed transcripts were forced aligned with speech recordings using a forced aligner \cite{yuan2008speaker}, which used the symbol “SIL” to denote pauses between words. Pauses at the beginning and end of the recordings, as well as transcripts from the interviewer, were removed. Following the method described in \cite{yuan2020disfluencies}, pauses were categorized into three groups: short (under 0.5 sec), medium (0.5-2 sec), and long (over 2 sec). These three bins of pauses were encoded using the punctuation marks “,”, “.”, and “. . . ”, respectively. Because all punctuations were removed from the processed transcripts, the aforementioned three encodings solely represented pause information. 

\vspace{-2mm}
\section{Experiments}
\vspace{-1mm}
\label{sec:Experiments}
We first describe the experimental setup in Section \ref{sec:Experimental Setup}, followed by the presentation and analysis of the experimental results in Section \ref{ref:Results and Analysis}. The results include the evaluation of the performance of four versions of two ASR models, as well as the AD detection results under the different conditions and methods. Overall, we conducted a series of comparative experiments, including four fine-tuned versions of two ASR models, PBFT versus TFT, manual transcripts versus ASR transcripts, the use versus non-use of pause encoding, and the inclusion versus exclusion of interviewer’s transcripts in the input sequence.

\vspace{-1mm}
\subsection{Experimental Setup}
\label{sec:Experimental Setup}
\vspace{-1mm}
We set maximum input length of BERT and RoBERTa to 512. We performed 10-fold cross-validation on the training set and tested on the test set. To enhance robustness, each AD detection system performed majority voting over the last three fine-tuning epochs of BERT or RoBERTa, and the two PLMs were also fused. For PBFT, the results of the templates in the two different positions were further integrated. Additionally, the evaluation of each system was repeated 15 times using 15 different random seeds. Mean, standard deviation (std) and the maximum among accuracy of all runs are used as performance measures. The following hyperparameters (slightly tuned) were chosen. For both fine-tuning methods, the number of training epochs was 20, the learning rate was 1e-05, AdamW optimizer was used, and the weight decay was 0.01. For TFT, the batch size was 4, while for PBFT, it was 1. All the parameters of both BERT and RoBERTa PLMs were fine-tuned. We converted the ASR outputs to lowercase, removed all punctuations, and used the manual transcripts as the ground truth for calculating the WER.
\vspace{-5mm}
\subsection{Results and Analysis}
\label{ref:Results and Analysis}
\vspace{-1mm}

\subsubsection{WER of the ASR models}
\vspace{-1mm}
The performance of four fine-tuned versions of the two ASR models (wav2vec 2.0 and Whisper) on the ADReSS dataset is shown in Table \ref{tab:WER of the ASR models}, which presents the mean WER results for five groups: All (training set plus test set), Healthy, Alzheimer, Training set, and Test set. The results indicate that the WER for each ASR model is higher in the Alzheimer group than in the Healthy group, which is expected, as AD patients tend to use fillers, incomplete words, and unclear articulation. The WERs for the Training set and Test set are approximately equal. The whisper-large-v3 model demonstrates the best performance across all ASR models: All (34.24\%), Healthy (28.47\%), Alzheimer (40.07\%), Training set (32.97\%), and Test set (37.06\%). Therefore, we selected its ASR transcripts for the subsequent experiments.

\begin{table*}[t!]
\vspace{-0.5cm}
\textit{}    \caption{AD detection results of TFT (the rows where the values of “Prompt Positions” are “-”) and PBFT (the remaining rows) under different conditions. Column 2 represents the use of either BERT or RoBERTa individually or the majority voting of their results. For PBFT, column 3 indicates whether a single prompt position (Before or After) is used, or the late fusion of both positions. Columns 4-6 display the results of 10-fold cross-validation on the training set, while columns 7-9 present the testing results on the test set. The results in each cell contain four specific sections, separated by “/”, corresponding to four different transcript inputs. These inputs, from left to right, are: “transcripts containing only the subjects”, “transcripts containing only the subjects and added pause encoding”, “transcripts containing both subjects and interviewers”, and “transcripts from the ASR model whisper-large-v3”. The results presented in parentheses are derived from the large-sized PLMs, whereas the others are from the base-sized PLMs.}
    \label{tab:results of TFT and PBFT}
    \centering
\resizebox{1.0\linewidth}{!}{
\begin{tabular}{|c|c|c|ccc|ccc|}
\hline
\multirow{2}{*}{Sys} & \multirow{2}{*}{PLMs}                                                       & \multirow{2}{*}{Prompt Positions} & \multicolumn{3}{c|}{Training Set CV Acc (\%)}                                                                                           & \multicolumn{3}{c|}{Test Set Acc (\%)}                                                                                                  \\ \cline{4-9} 
                     &                                                                             &                                   & \multicolumn{1}{c|}{Mean}                            & \multicolumn{1}{c|}{Std}                       & Maximum                         & \multicolumn{1}{c|}{Mean}                            & \multicolumn{1}{c|}{Std}                       & Maximum                         \\ \hline
1                    & \multirow{4}{*}{BERT}                                                       & -                                 & \multicolumn{1}{c|}{80.4/80.8/78.9/75.5}             & \multicolumn{1}{c|}{2.1/2.7/3.0/1.4}           & 82.3/83.7/81.9/78.5& \multicolumn{1}{c|}{81.6/82.4/79.6/70.8}             & \multicolumn{1}{c|}{3.0/1.3/2.3/2.9}           & 84.6/84.3/83.3/76.4\\ \cline{1-1} \cline{3-9} 
2                    &                                                                             & Before                            & \multicolumn{1}{c|}{81.2(81.7)/81.5(83.3)/82.4/75.7} & \multicolumn{1}{c|}{1.8(2.0)/2.3(2.5)/1.8/1.6} & 86.1(85.2)/86.1(86.1)/86.1/78.7 & \multicolumn{1}{c|}{79.9(86.0)/84.4(84.6)/80.0/72.1} & \multicolumn{1}{c|}{2.2(3.0)/3.8(2.5)/3.0/2.6} & 83.3(91.7)/87.5(91.7)/85.4/77.1 \\ \cline{1-1} \cline{3-9} 
3                    &                                                                             & After                             & \multicolumn{1}{c|}{74.0(81.5)/74.5(80.2)/73.6/70.9} & \multicolumn{1}{c|}{2.2(1.9)/2.8(3.4)/3.1/2.2} & 78.7(82.4)/80.6(83.3)/77.8/75.0 & \multicolumn{1}{c|}{80.1(81.3)/80.7(84.0)/77.8/66.9} & \multicolumn{1}{c|}{3.3(4.5)/4.5(4.7)/2.9/2.6} & 85.4(93.8)/85.4(93.8)/81.3/70.8 \\ \cline{1-1} \cline{3-9} 
4                    &                                                                             & Before+After                      & \multicolumn{1}{c|}{79.8(82.1)/82.2(82.8)/78.4/74.9} & \multicolumn{1}{c|}{2.0(2.0)/2.8(2.0)/1.8/2.1} & 83.3(87.0)/83.3(87.0)/82.4/77.8 & \multicolumn{1}{c|}{83.6(86.5)/84.2(86.5)/77.2/70.8} & \multicolumn{1}{c|}{2.5(2.0)/2.7(3.4)/2.8/2.3} & 87.5(89.6)/87.5(91.7)/81.3/75.0 \\ \hline
5                    & \multirow{4}{*}{RoBERTa}                                                    & -                                 & \multicolumn{1}{c|}{81.3/82.4/80.0/77.3}             & \multicolumn{1}{c|}{2.2/2.3/2.4/1.9}           & 83.6/85.2/82.2/78.9& \multicolumn{1}{c|}{82.1/82.9/80.6/74.4}             & \multicolumn{1}{c|}{2.4/1.7/2.9/3.3}           & 85.4/85.4/85.4/75.2\\ \cline{1-1} \cline{3-9} 
6                    &                                                                             & Before                            & \multicolumn{1}{c|}{81.4(82.0)/83.7(83.3)/80.2/76.7} & \multicolumn{1}{c|}{1.4(1.9)/2.0(4.1)/1.5/1.7} & 84.3(84.3)/87.0(88.0)/83.3/80.6 & \multicolumn{1}{c|}{80.3(82.9)/80.1(84.2)/77.5/74.6} & \multicolumn{1}{c|}{2.5(3.8)/3.7(4.6)/2.8/2.0} & 85.4(87.5)/85.4(91.8)/81.3/79.2\\ \cline{1-1} \cline{3-9} 
7                    &                                                                             & After                             & \multicolumn{1}{c|}{80.2(82.5)/83.3(82.0)/75.7/74.1} & \multicolumn{1}{c|}{1.8(2.0)/1.8(2.6)/1.7/2.0} & 84.3(87.0)/86.1(84.3)/78.7/78.7 & \multicolumn{1}{c|}{80.7(82.5)/83.2(82.8)/80.7/70.0} & \multicolumn{1}{c|}{2.3(3.4)/4.2(3.9)/3.4/4.4} & 85.4(87.5)/91.7(91.7)/85.4/79.2\\ \cline{1-1} \cline{3-9} 
8                    &                                                                             & Before+After                      & \multicolumn{1}{c|}{83.0(84.1)/84.0(84.2)/80.1/76.8} & \multicolumn{1}{c|}{1.7(1.2)/1.5(3.2)/1.2/2.8} & 86.1(86.1)/86.1(84.3)/83.3/79.6 & \multicolumn{1}{c|}{82.9(83.2)/84.7(84.9)/79.6/72.9} & \multicolumn{1}{c|}{2.2(3.9)/3.4(5.8)/3.2/3.3} & 87.5(89.6)/89.0(93.8)/85.4/81.3 \\ \hline
9                    & \multirow{4}{*}{\begin{tabular}[c]{@{}c@{}}BERT\\ +\\ RoBERTa\end{tabular}} & -                                 & \multicolumn{1}{c|}{82.8/83.3/81.6/77.6}             & \multicolumn{1}{c|}{2.1/2.2/2.2/2.1}           & 84.2/86.7/82.8/82.4& \multicolumn{1}{c|}{83.6/84.2/80.4/72.6}             & \multicolumn{1}{c|}{2.5/1.6/2.5/1.8}           & 87.5/91.7/85.4/77.1\\ \cline{1-1} \cline{3-9} 
10                   &                                                                             & Before                            & \multicolumn{1}{c|}{81.3(81.3)/82.7(82.5)/82.1/78.4} & \multicolumn{1}{c|}{1.3(1.3)/2.4(2.4)/1.4/1.8} & 85.2(85.2)/88.0(88.0)/86.1/84.3 & \multicolumn{1}{c|}{82.5(84.0)/82.9(84.0)/76.9/72.8} & \multicolumn{1}{c|}{2.5(3.0)/4.2(2.6)/2.7/1.6} & 89.6(91.7)/91.7(91.7)/83.3/79.2 \\ \cline{1-1} \cline{3-9} 
11                   &                                                                             & After                             & \multicolumn{1}{c|}{78.5(78.5)/79.4(79.4)/75.2/76.5} & \multicolumn{1}{c|}{2.1(2.1)/1.7(1.7)/2.1/2.3} & 83.3(83.3)/84.3(84.3)/80.6/81.5 & \multicolumn{1}{c|}{83.5(84.7)/82.6(82.6)/78.9/68.9} & \multicolumn{1}{c|}{3.1(3.9)/3.4(4.0)/2.8/2.4} & 91.7(93.8)/89.6(93.8)/87.5/75.0 \\ \cline{1-1} \cline{3-9} 
12                   &                                                                             & Before+After                      & \multicolumn{1}{c|}{83.1(83.1)/84.2(84.2)/81.1/77.6} & \multicolumn{1}{c|}{1.2(1.2)/1.2(1.3)/1.6/1.5} & 86.1(86.1)/88.0(88.0)/85.2/82.4 & \multicolumn{1}{c|}{85.0(85.0)/86.5(\textbf{87.9})/79.7/72.6} & \multicolumn{1}{c|}{2.2(2.6)/2.6(\textbf{3.3})/2.6/2.4} & 91.7(89.6)/91.7(\textbf{95.8})/85.4/79.2 \\ \hline
\end{tabular}}
\vspace{-0.6cm}
\end{table*}

\vspace{-2mm}
\subsubsection{AD detection results}
\vspace{-1mm}
The performance measures of TFT and PBFT for AD detection under different conditions are shown in Table \ref{tab:results of TFT and PBFT}, which reveals several main trends. Firstly, PBFT generally outperforms TFT in terms of the mean or maximum accuracy, indicating that PBFT is more effective for AD detection. For PBFT, there is no significant performance difference between prompt templates at different positions. Late fusion of templates from two different positions within any individual PLMs does not lead to performance improvement, likely due to insufficient diversity between the different position templates among the resulting PLMs, resulting in a lack of complementarity. However, majority voting between different PLMs (BERT + RoBERTa) across different prompt positions can enhance detection performance and improve stability (smaller std). This may because the late fusion of the two PLMs can further leverage their complementarity, since RoBERTa's performance is generally, but not consistently, superior to BERT.

\begin{table}[b]
\vspace{-0.6cm}
  \caption{Comparison of the best test set accuracy presented in this paper with the recent SOTA results based on different modalities for the ADReSS dataset.}
    \label{tab:Comparison of SOTA}
    \centering
\resizebox{1.0\linewidth}{!}{
\begin{tabular}{ccc}
\toprule
Literature                                  & Modality                                                                                                     & Test Set Acc (\%)                                             \\
\midrule
Laguarta et al.\cite{laguarta2021longitudinal} & Audio + Pre-trained Biomarkers& 93.8                                                          \\
\midrule
Martinc et al. \cite{martinc2021temporal}      & Audio + Manual Transcript& 93.8                                                          \\
\midrule
Wang et al. \cite{wang2022exploring}           & Manual Transcript& 91.7\\
\midrule
Our work                                    & Manual Transcript& \textbf{95.8}\\
\bottomrule
\end{tabular}
}
\end{table}

Regarding results from different transcript inputs, TFT and PBFT reveal consistent trends. Results from transcripts containing only the subjects outperform those including the interviewers, suggesting that interviewers’ transcripts may interfere with the model's extraction of linguistic features relevant to distinguishing AD from non-AD. The addition of pause encoding to the manual transcripts consistently enhances performance, indicating that it aids the model in capturing features such as disfluencies associated with AD. Although we selected the relatively well-performing Whisper model, its ASR-based transcripts reduce AD detection performance compared to manual transcripts. This may be due to the difficulty in extracting useful classification features from ASR-based transcripts with high WER. Recent studies \cite{LI2024104598,li2024speech} have indicated that imperfect ASR transcripts can sometimes provide valuable cues for downstream tasks (e.g., AD detection, emotion recognition), thereby improving the performance of the corresponding tasks. The discrepancy in conclusions in this study may be attributed to the different ASR models and methods used. This paper lacks sufficient investigation into the impact of ASR transcripts on downstream task performance (as it is not the main focus of this paper), necessitating further research for a more in-depth analysis in this area.
%

To investigate whether large-sized PLMs provide improvements over base-sized ones, we conducted PBFT with large-sized PLMs. These results are presented in parentheses in Table  \ref{tab:results of TFT and PBFT}. As noted in previous analysis, including interviewer transcripts will reduce AD detection performance, and the performance of ASR-based transcripts with errors is lower than that of manual transcripts. Therefore, we only used two types of inputs in this experiment: transcripts containing only the subjects (left side of the first “/”) and transcripts containing only the subjects and added pause encoding (right side of the first “/”). Comparing these results with those from the base-sized PLMs, we can observe that increasing the model size slightly improves AD detection performance. Ultimately, we achieved a maximum AD detection accuracy of 95.8\% on the test set when inputs are manual transcripts containing only the subjects and added pause encoding (with mean 87.9\%, std 3.3\%) (the bolded values in Sys. 12), reaching SOTA results using only manual transcripts. 

The best AD detection performance presented in this paper were further compared in Table \ref{tab:Comparison of SOTA} with SOTA results reported in recent literature using the same ADReSS dataset to demonstrate their competitiveness.

\vspace{-3mm}
\section{Conclusions}
\label{sec:Conclusions}
\vspace{-1mm}
In this paper, we proposed a method for AD detection by combining prompt-based fine-tuning of PLMs with pause encoding. Majority voting and ensemble techniques were further applied across BERT and RoBERTa using different random seeds. Experimental results demonstrated the superiority of prompt-based fine-tuning over traditional fine-tuning, with pause encoding enhancing AD detection performance in both methods. To compare with manual transcripts, we employed the ASR model whisper-large-v3, which demonstrated relatively better performance, to obtain ASR-based transcripts. We found that ASR transcripts with hign WER reduce AD detection performance compared to manual transcripts. We also investigated the impact of interviewer’s transcripts on AD detection performance and found that they interfere with the model's ability to correctly distinguish between AD and non-AD.  Ultimately, maximum detection accuracy of 95.8\% (with mean 87.9\%, std 3.3\%) was obtained using manual transcripts, achieving SOTA performance in AD detection using only transcripts on the ADReSS test set. We hope that the research presented in this paper will provide valuable insights for the development of more effective automatic AD detection techniques. Moving forward, we aim to investigate the relationship between ASR errors and linguistic features extracted from the speech of AD patients to better understand what the model has learned.

\bibliographystyle{IEEEtran}

\bibliography{mybib}

\begin{thebibliography}{10}
\providecommand{\url}[1]{#1}
\csname url@samestyle\endcsname
\providecommand{\newblock}{\relax}
\providecommand{\bibinfo}[2]{#2}
\providecommand{\BIBentrySTDinterwordspacing}{\spaceskip=0pt\relax}
\providecommand{\BIBentryALTinterwordstretchfactor}{4}
\providecommand{\BIBentryALTinterwordspacing}{\spaceskip=\fontdimen2\font plus
\BIBentryALTinterwordstretchfactor\fontdimen3\font minus \fontdimen4\font\relax}
\providecommand{\BIBforeignlanguage}[2]{{%
\expandafter\ifx\csname l@#1\endcsname\relax
\typeout{** WARNING: IEEEtran.bst: No hyphenation pattern has been}%
\typeout{** loaded for the language `#1'. Using the pattern for}%
\typeout{** the default language instead.}%
\else
\language=\csname l@#1\endcsname
\fi
#2}}
\providecommand{\BIBdecl}{\relax}
\BIBdecl

\bibitem{nestor2004advances}
P.~J. Nestor, P.~Scheltens, and J.~R. Hodges, ``Advances in the early detection of alzheimer's disease,'' \emph{Nature medicine}, vol.~10, no. Suppl 7, pp. S34--S41, 2004.

\bibitem{de2019brief}
E.~E. De~Roeck, P.~P. De~Deyn, E.~Dierckx, and S.~Engelborghs, ``Brief cognitive screening instruments for early detection of alzheimer’s disease: a systematic review,'' \emph{Alzheimer's research \& therapy}, vol.~11, pp. 1--14, 2019.

\bibitem{luz2020fuente}
S.~Luz and F.~Haide, ``Fuente s. dl, fromm d, macwhinney b,'' \emph{Alzheimer's Dementia Recognition through Spontaneous Speech: The ADReSS Challenge. Proc Interspeech}, pp. 2172--6, 2020.

\bibitem{luz2021detecting}
S.~Luz, F.~Haider, S.~de~la Fuente, D.~Fromm, and B.~MacWhinney, ``Detecting cognitive decline using speech only: The adresso challenge,'' \emph{arXiv preprint arXiv:2104.09356}, 2021.

\bibitem{pan2021using}
Y.~Pan, B.~Mirheidari, J.~M. Harris, J.~C. Thompson, M.~Jones, J.~S. Snowden, D.~Blackburn, and H.~Christensen, ``Using the outputs of different automatic speech recognition paradigms for acoustic-and bert-based alzheimer's dementia detection through spontaneous speech.'' in \emph{Interspeech}, 2021, pp. 3810--3814.

\bibitem{liu2024clever}
Y.-L. Liu, R.~Feng, J.-H. Yuan, and Z.-H. Ling, ``Clever hans effect found in automatic detection of alzheimer's disease through speech,'' \emph{arXiv preprint arXiv:2406.07410}, 2024.

\bibitem{mei2023ustc}
K.~Mei, X.~Ding, Y.~Liu, Z.~Guo, F.~Xu, X.~Li, T.~Naren, J.~Yuan, and Z.~Ling, ``The ustc system for adress-m challenge,'' in \emph{ICASSP 2023-2023 IEEE International Conference on Acoustics, Speech and Signal Processing (ICASSP)}.\hskip 1em plus 0.5em minus 0.4em\relax IEEE, 2023, pp. 1--2.

\bibitem{syed2020automated}
M.~S.~S. Syed, Z.~S. Syed, M.~Lech, and E.~Pirogova, ``Automated screening for alzheimer's dementia through spontaneous speech.'' in \emph{Interspeech}, vol. 2020, 2020, pp. 2222--6.

\bibitem{syed2021tackling}
Z.~S. Syed, M.~S.~S. Syed, M.~Lech, and E.~Pirogova, ``Tackling the adresso challenge 2021: The muet-rmit system for alzheimer's dementia recognition from spontaneous speech.'' in \emph{Interspeech}, 2021, pp. 3815--3819.

\bibitem{rohanian2021alzheimer}
M.~Rohanian, J.~Hough, and M.~Purver, ``Alzheimer's dementia recognition using acoustic, lexical, disfluency and speech pause features robust to noisy inputs,'' \emph{arXiv preprint arXiv:2106.15684}, 2021.

\bibitem{martinc2020tackling}
M.~Martinc and S.~Pollak, ``Tackling the adress challenge: A multimodal approach to the automated recognition of alzheimer's dementia.'' in \emph{Interspeech}, 2020, pp. 2157--2161.

\bibitem{qiao2021alzheimer}
Y.~Qiao, X.~Yin, D.~Wiechmann, and E.~Kerz, ``Alzheimer's disease detection from spontaneous speech through combining linguistic complexity and (dis) fluency features with pretrained language models,'' \emph{arXiv preprint arXiv:2106.08689}, 2021.

\bibitem{santos2017enriching}
L.~B.~d. Santos, E.~A. Corr{\^e}a~Jr, O.~N. Oliveira~Jr, D.~R. Amancio, L.~L. Mansur, and S.~M. Alu{\'\i}sio, ``Enriching complex networks with word embeddings for detecting mild cognitive impairment from speech transcripts,'' \emph{arXiv preprint arXiv:1704.08088}, 2017.

\bibitem{li2021comparative}
J.~Li, J.~Yu, Z.~Ye, S.~Wong, M.~Mak, B.~Mak, X.~Liu, and H.~Meng, ``A comparative study of acoustic and linguistic features classification for alzheimer's disease detection,'' in \emph{ICASSP 2021-2021 IEEE International Conference on Acoustics, Speech and Signal Processing (ICASSP)}.\hskip 1em plus 0.5em minus 0.4em\relax IEEE, 2021, pp. 6423--6427.

\bibitem{wang2022exploring}
Y.~Wang, T.~Wang, Z.~Ye, L.~Meng, S.~Hu, X.~Wu, X.~Liu, and H.~Meng, ``Exploring linguistic feature and model combination for speech recognition based automatic ad detection,'' \emph{arXiv preprint arXiv:2206.13758}, 2022.

\bibitem{zhu2021wavbert}
Y.~Zhu, A.~Obyat, X.~Liang, J.~A. Batsis, and R.~M. Roth, ``Wavbert: Exploiting semantic and non-semantic speech using wav2vec and bert for dementia detection,'' in \emph{Interspeech}, vol. 2021.\hskip 1em plus 0.5em minus 0.4em\relax NIH Public Access, 2021, p. 3790.

\bibitem{wang2023exploiting}
Y.~Wang, J.~Deng, T.~Wang, B.~Zheng, S.~Hu, X.~Liu, and H.~Meng, ``Exploiting prompt learning with pre-trained language models for alzheimer’s disease detection,'' in \emph{ICASSP 2023-2023 IEEE International Conference on Acoustics, Speech and Signal Processing (ICASSP)}.\hskip 1em plus 0.5em minus 0.4em\relax IEEE, 2023, pp. 1--5.

\bibitem{devlin2018bert}
J.~Devlin, M.-W. Chang, K.~Lee, and K.~Toutanova, ``Bert: Pre-training of deep bidirectional transformers for language understanding,'' \emph{arXiv preprint arXiv:1810.04805}, 2018.

\bibitem{liu2019roberta}
Y.~Liu, M.~Ott, N.~Goyal, J.~Du, M.~Joshi, D.~Chen, O.~Levy, M.~Lewis, L.~Zettlemoyer, and V.~Stoyanov, ``Roberta: A robustly optimized bert pretraining approach,'' \emph{arXiv preprint arXiv:1907.11692}, 2019.

\bibitem{zhu2023prompt}
Y.~Zhu, Y.~Wang, J.~Qiang, and X.~Wu, ``Prompt-learning for short text classification,'' \emph{IEEE Transactions on Knowledge and Data Engineering}, 2023.

\bibitem{feng2024prompt}
K.~Feng, L.~Huang, K.~Wang, W.~Wei, and R.~Zhang, ``Prompt-based learning framework for zero-shot cross-lingual text classification,'' \emph{Engineering Applications of Artificial Intelligence}, vol. 133, p. 108481, 2024.

\bibitem{xu2022match}
S.~Xu, L.~Pang, H.~Shen, and X.~Cheng, ``Match-prompt: Improving multi-task generalization ability for neural text matching via prompt learning,'' in \emph{Proceedings of the 31st ACM International Conference on Information \& Knowledge Management}, 2022, pp. 2290--2300.

\bibitem{zheng2024exploring}
F.~Zheng, J.~Cao, W.~Yu, Z.~Chen, N.~Xiao, and Y.~Lu, ``Exploring low-resource medical image classification with weakly supervised prompt learning,'' \emph{Pattern Recognition}, vol. 149, p. 110250, 2024.

\bibitem{baevski2020wav2vec}
A.~Baevski, Y.~Zhou, A.~Mohamed, and M.~Auli, ``wav2vec 2.0: A framework for self-supervised learning of speech representations,'' \emph{Advances in neural information processing systems}, vol.~33, pp. 12\,449--12\,460, 2020.

\bibitem{radford2023robust}
A.~Radford, J.~W. Kim, T.~Xu, G.~Brockman, C.~McLeavey, and I.~Sutskever, ``Robust speech recognition via large-scale weak supervision,'' in \emph{International Conference on Machine Learning}.\hskip 1em plus 0.5em minus 0.4em\relax PMLR, 2023, pp. 28\,492--28\,518.

\bibitem{becker1994natural}
J.~T. Becker, F.~Boiler, O.~L. Lopez, J.~Saxton, and K.~L. McGonigle, ``The natural history of alzheimer's disease: description of study cohort and accuracy of diagnosis,'' \emph{Archives of neurology}, vol.~51, no.~6, pp. 585--594, 1994.

\bibitem{macwhinney2014childes}
B.~MacWhinney, \emph{The CHILDES project: Tools for analyzing talk, Volume I: Transcription format and programs}.\hskip 1em plus 0.5em minus 0.4em\relax Psychology Press, 2014.

\bibitem{ding2021openprompt}
N.~Ding, S.~Hu, W.~Zhao, Y.~Chen, Z.~Liu, H.-T. Zheng, and M.~Sun, ``Openprompt: An open-source framework for prompt-learning,'' \emph{arXiv preprint arXiv:2111.01998}, 2021.

\bibitem{yuan2008speaker}
J.~Yuan, M.~Liberman \emph{et~al.}, ``Speaker identification on the scotus corpus,'' \emph{Journal of the Acoustical Society of America}, vol. 123, no.~5, p. 3878, 2008.

\bibitem{yuan2020disfluencies}
J.~Yuan, Y.~Bian, X.~Cai, J.~Huang, Z.~Ye, and K.~Church, ``Disfluencies and fine-tuning pre-trained language models for detection of alzheimer's disease.'' in \emph{Interspeech}, vol. 2020, 2020, pp. 2162--6.

\bibitem{laguarta2021longitudinal}
J.~Laguarta and B.~Subirana, ``Longitudinal speech biomarkers for automated alzheimer's detection,'' \emph{frontiers in Computer Science}, vol.~3, p. 624694, 2021.

\bibitem{martinc2021temporal}
M.~Martinc, F.~Haider, S.~Pollak, and S.~Luz, ``Temporal integration of text transcripts and acoustic features for alzheimer's diagnosis based on spontaneous speech,'' \emph{Frontiers in Aging Neuroscience}, vol.~13, p. 642647, 2021.

\bibitem{LI2024104598}
C.~Li, W.~Xu, T.~Cohen, and S.~Pakhomov, ``Useful blunders: Can automated speech recognition errors improve downstream dementia classification?'' \emph{Journal of Biomedical Informatics}, vol. 150, p. 104598, 2024.

\bibitem{li2024speech}
Y.~Li, P.~Bell, and C.~Lai, ``Speech emotion recognition with asr transcripts: A comprehensive study on word error rate and fusion techniques,'' \emph{arXiv preprint arXiv:2406.08353}, 2024.

\end{thebibliography}

\end{document}